\documentclass[11pt,a4paper]{article}
\usepackage{jcappub}
\usepackage{soul} 
\usepackage[normalem]{ulem}

\title{Gravity in mimetic scalar-tensor theories after GW170817}
\author[a,b]{Alexander Ganz,}
\author[a,b,c]{Nicola Bartolo,}
\author[a]{Purnendu Karmakar,}
\author[a,b,c,d]{Sabino Matarrese}
\affiliation[a]{ Dipartimento di Fisica e Astronomia ``G. Galilei", \\
	Universit\`a degli Studi di Padova, via Marzolo 8, I-35131 Padova, Italy}
\affiliation[b]{INFN, Sezione di Padova, \\ via Marzolo 8, I-35131 Padova, Italy}
\affiliation[c]{INAF - Osservatorio Astronomico di Padova, \\ Vicolo dell’Osservatorio 5, I-35122 Padova, Italy}
\affiliation[d]{Gran Sasso Science Institute, \\ Viale F. Crispi 7, I-67100 L'Aquila, Italy}
\emailAdd{alexander.ganz@pd.infn.it}
\emailAdd{nicola.bartolo@pd.infn.it}
\emailAdd{purnendu.karmakar@pd.infn.it}
\emailAdd{sabino.matarrese@pd.infn.it} 

\abstract{We derive the most general mimetic scalar-tensor theory assuming a healthy ``seed" action and accounting for the constraints on the speed of gravitational-wave propagation arising 
from the GW170817 event. By analysing linear perturbations around a flat FLRW background in this model, we obtain a suitable form of the Poisson equation, which allows us to 
calculate the effective gravitational constant felt by ``ordinary" matter. By restricting to a minimally coupled model, such an effective gravitational 
constant is equivalent to that obtained within General Relativity, with cold dark matter plus a perfect fluid dark energy component, 
with vanishing sound speed. Assuming, further, a $\Lambda$CDM background, the effective gravitational constant cannot be distinguished from that of the standard $\Lambda$CDM model, at 
linear order. For the full non-minimally coupled mimetic gravity model we obtain a non-vanishing gravitational slip and an effective gravitational constant which always differs from that of standard 
$\Lambda$CDM. }

\arxivnumber{}

\newcommand{\h}{\mathcal{H}}
\newcommand{\md}{\mathrm{d}}

\notoc 
\begin{document}
	\maketitle
	\flushbottom
	
	\section{Introduction}
	In \cite{Chamseddine:2013kea} mimetic matter (aka ``mimetic gravity") was introduced as a modification of General Relativity (GR). By performing a non-invertible conformal transformation of the 
	Einstein-Hilbert action
	\begin{align}
	g_{\mu\nu} = - \left(\tilde g^{\alpha\beta}\partial_\alpha \varphi \partial_\beta \varphi\right) \tilde g_{\mu\nu}.
	\end{align}
	one obtains a conformally invariant theory where the additional degree of freedom can mimic a cold dark matter contribution. Alternatively, by gauge fixing the conformal symmetry \cite{Barvinsky:2013mea,Chaichian:2014qba}, the mimetic gravity model can be equivalently derived by adding the mimetic constraint via a Lagrange multiplier to the original seed theory into the action $\lambda 
	\left(g^{\mu\nu}\partial_\mu \varphi \partial_\nu \varphi +1\right)$ \cite{Golovnev:2013jxa}.
	The idea of a non-invertible transformation has been extended to a broad class of different modified gravity models  as scalar-tensor theories \cite{Lim:2010yk,Chamseddine:2014vna,Arroja:2015wpa,Arroja:2015yvd,Arroja:2017msd,Langlois:2018jdg,Takahashi:2017pje}, massive gravity \cite{Chamseddine:2018gqh}, $f(R)$ gravity \cite{Nojiri:2014zqa,Odintsov:2015cwa,Odintsov:2015wwp}   (see \cite{Sebastiani:2016ras} for a review).
	
	A necessary requirement for a modification of General Relativity (GR) is that the introduction of higher derivative terms does not introduce any ghost degree of freedom. 
	Due to the Ostrogradski theorem (see for instance \cite{Woodard:2015zca}) this leads to strict conditions on the possible higher derivative terms. For a general scalar-tensor theory, higher 
	derivative terms of the scalar field are allowed if specific degeneration conditions are fulfilled, which ensure that there is no additional Ostrogradski ghost degree of freedom \cite{Langlois:2015cwa,Crisostomi:2016czh,Achour:2016rkg,BenAchour:2016fzp}. This class of theories are called "Degenerate Higher Order Scalar-Tensor" (DHOST) theories. Mimetic DHOST theories are 
	discussed in \cite{Langlois:2018jdg}.
	
	Recently, the detection of the gravitational wave (GW) event, GW170817 \cite{TheLIGOScientific:2017qsa}, due to a binary neutron star merging and its associated gamma-ray burst (GRB) 
	signal, GRB170817A \cite{Monitor:2017mdv}, provided tight constraints on the speed of gravitational wave propagation $\vert c_t^2/c^2-1\vert \le 5\times 10^{-16}$. This has been used to 
	constrain and exclude several modified gravity models that predict a speed of gravitational waves different from the speed of light \cite{Nishizawa:2014zna, Lombriser:2015sxa, Ezquiaga:2017ekz, Lombriser:2016yzn, Bettoni:2016mij, Sakstein:2017xjx,Creminelli:2017sry,Baker:2017hug, Jain:2015edg, Arai:2017hxj,Heisenberg:2017qka}.
	In \cite{Langlois:2017dyl,Crisostomi:2017lbg, Bartolo:2017ibw,Dima:2017pwp} the constraints due to the speed of the gravitational waves have been derived for the DHOST models.
	
	In \cite{Arroja:2015yvd,Arroja:2017msd,Arroja:2015wpa} a mimetic Horndeski model is introduced. Restricting to a cubic mimetic Horndeski model a linear analysis around the Friedman-Lema\^{i}tre-Robertson-
	Walker (FLRW) background shows that the evolution equation for the scalar gravitational perturbation is equivalent to that in GR with cold dark matter (CDM) and perfect fluid dark energy (PFDE), 
	with vanishing sound speed, thus providing a natural implementation of a unified dark matter-dark energy model (see e.g. \cite{Bertacca:2010ct}, for a review). By considering a $\Lambda$CDM 
	background, the cubic mimetic Horndeski model cannot be distinguished from the standard $\Lambda$CDM model up to first order.
	
	The Poisson equation and the gravitational slip are commonly used in the literature to test GR and constrain modified gravity models \cite{Zhang:2007nk,Amendola:2007rr,Jain:2007yk,Bertschinger:2008zb,Capela:2014xta}. The gravitational slip measures the deviation between the Newtonian $\Phi$ and the longitudinal $\Psi$ gravitational potentials. In the absence of anisotropic stress they are equivalent for GR, while they can differ in modified gravity models, as for instance in many non-minimally coupled scalar-tensor theories \cite{Bellini:2014fua}. The Poisson equation deviations from GR are described by an effective gravitational constant which can be explicitly time- and scale-dependent, in contrast to GR.
	
	In the first part of this paper, section \ref{sec:Mimetic_scalar-tensor_theory} and \ref{sec:Constraints_from_GW}, we derive the most general mimetic scalar-tensor theory, accounting for the constraints imposed on the speed of gravitational waves and for the restriction of a healthy seed theory.\footnote{By this we mean that the original seed theory does not suffer from instabilities due to higher-order derivative terms.} In the following section \ref{sec:Poisson_Equation_for_General_Relativity} we review the derivation of the Poisson equation in standard GR and describe an alternative derivation which is used to obtain the Poisson equation for GR, with CDM and PFDE. Further, in section \ref{sec:Mimetic_Gravity} we adopt the new method to derive the Poisson equation for the most general mimetic scalar-tensor theory discussed before. Using this result we analyze the case of (non)-minimal coupling of the mimetic scalar field with the curvature separately in more detail. We also provide an expression for the gravitational slip. Finally, in section \ref{sec:Conclusions} we give a discussion of our main results. 
	
	In this paper we adopt units such that the speed of light and the reduced Planck mass are unity. Further, we adopt the $(-,+,+,+)$ signature for the metric. Greek indices run from 0 to 3 and latin ones from 1 to 3.	
	
	\section{Mimetic scalar-tensor theory}
	\label{sec:Mimetic_scalar-tensor_theory}
	The action for the general scalar-tensor theory up to cubic dependence on the second-derivatives of the scalar field can be written as
	\cite{Crisostomi:2016czh,Achour:2016rkg,BenAchour:2016fzp,Motohashi:2016ftl}
	\begin{align}
	S &= \int\md^4x\,\sqrt{-g}\, \left( f_2(\varphi,X) R + f_3(\varphi,X) G_{\mu\nu} \varphi^\mu \varphi^\nu +  K(\varphi,X) + G(\varphi,X) \Box\varphi\right) \nonumber \\ &+ \int\md^4x\,\sqrt{-g}\,
	\left( \sum_i a_i(\varphi,X) L^{(2)}_{i} + \sum_j b_j(\varphi,X) L_{j}^{(3)}\right) 
	\label{a:dhost}
	\end{align}
 	where $X=g^{\mu\nu}\varphi_\mu\varphi_\nu$, $\nabla_\mu\varphi \equiv \varphi_\mu$ and $a_i$ and $b_i$ are functions of $\varphi$ and $X$. Further, $L^{(2)}$ or $L^{(3)}$ are all possible 
	contractions of a scalar field up to second or third polynomial degree in second-order derivatives of the scalar field \cite{BenAchour:2016fzp}. 
	Together with the degeneracy conditions, which relates the parameter $a_i$ and $b_j$, the action (\ref{a:dhost}) defines the DHOST theory \cite{Crisostomi:2016czh,Achour:2016rkg,BenAchour:2016fzp}. Higher than third degree terms which could be in principle added to the action above are ruled out by the speed of gravitational waves constraint \cite{Langlois:2017mxy}. Indeed, the Horndeski action is the subclass of the above DHOST action defined in \cite{BenAchour:2016fzp}.
	
The action above in Eq. (\ref{a:dhost}) will be assumed as the ``seed action" for our mimetic scalar-tensor theory, whose action can be written as 
	\begin{equation}
	S_{mimetic} \equiv S_{seed} - \int\md^4x\,\sqrt{-g}\,\lambda \left(X + 1\right) \;, \label{a:m:dhost}
	\end{equation}
	In general, the mimetic constraint could be generalized to $\lambda \left(b(\varphi)X+1\right)$. However, the Lagrangian is invariant under field redefinitions and therefore one can reabsorb 
	$b(\varphi)$ through a scalar field redefinition, as long as $b(\varphi)$ is integrable (see the discussion in \cite{Arroja:2015wpa,Takahashi:2017pje,Langlois:2018jdg,Arroja:2015yvd}). 

	Due to the mimetic constraint, the $X$ dependence of the functions can be neglected, since it can always be reabsorbed into a re-defined Lagrange multiplier, see \cite{Langlois:2018jdg,MimeticHamiltonian}
	Furthermore, any term containing higher-order derivatives of $\varphi$ in $L^{(2)}$ or $L^{(3)}$, having the form
	\begin{equation}
	\theta_n \equiv \varphi^\mu\varphi^\nu \left[\varphi\right]_{\mu\nu}^n \;,
	\end{equation}
	will vanish identically, since it involves covariant derivatives of $X$. We use the notation $\theta_1= \varphi^\mu \varphi^\nu \varphi_{\mu\nu}$, $\theta_2 = \varphi_\alpha 
	\varphi_{\mu\nu} \varphi^{\alpha\mu} \varphi^\nu$. Therefore, the functions $a_3$, $a_4$ ,$a_5$, $b_4$,...,$b_{10}$ can be set to zero without loss of generality. The action can be finally written as 
	\begin{align}
	S &= \int\md^4x\,\sqrt{-g}\, \left( f_2 R + f_3\, G_{\mu\nu} \varphi^\mu \varphi^\nu +  K + G\, \Box\varphi+ a_1\, \varphi^{\mu\nu}\varphi_{\mu\nu} + a_2 \left(\Box \varphi\right)^2  \right) \nonumber \\ 
	&+ \int\md^4x\,\sqrt{-g}\, \left( b_1 \left(\Box\varphi\right)^3 + b_2\,\Box\varphi \varphi^{\mu\nu}\varphi_{\mu\nu} + b_3\, 
	\varphi^{\mu\nu}\varphi_{\rho\nu} \varphi^\rho_\mu \right) - \int\md^4x\,\sqrt{-g}\,\lambda \left(X + 1\right).
	\end{align}
	All the functions depend now only on $\varphi$. 
	
	As a next step one can further simplify the action. In \cite{Langlois:2018jdg} it is discussed that one can set $f_3=0$, without any loss of generality, by redefining the functions $f_2,a_1,a_2$ and 
	$G$. Further, since $G$ is only a function of $\varphi$ one can partially integrate this part, which yields 
	\begin{equation}
	\int\md^4x\,\sqrt{-g}\, \left(K(\varphi) + G(\varphi)\,\Box\varphi\right) = \int\md^4x\,\sqrt{-g}\, \left( K(\varphi) - G_{\varphi} X \right) \equiv \int\md^4x\,\sqrt{-g}\, V(\varphi,X),
	\end{equation}
	where we use the notation $G_\varphi \equiv \partial_\varphi G$.
	Once again, one can insert the mimetic constraint $X=-1$ to obtain a general potential 
	$V(\varphi)= K(\varphi) + G_\varphi(\varphi)$ \footnote{Notice that the sign of $V(\varphi)$ here is opposite to the one commonly used for a potential.}, which would read $V=K-G_{3,\varphi}$ in the framework of Horndeski (see \cite{Arroja:2015wpa} for the sign convention of the 
	Horndeski action). Finally, one obtains
	\begin{align}
	S &= \int\md^4x\,\sqrt{-g}\, \left( f_2 R + V+ a_1\, \varphi^{\mu\nu}\varphi_{\mu\nu} + a_2 \left(\Box \varphi\right)^2  +  b_1 \left(\Box\varphi\right)^3 + b_2\,\Box\varphi \varphi^{\mu\nu}
	\varphi_{\mu\nu} + b_3\, \varphi^{\mu\nu}\varphi_{\rho\nu} \varphi^\rho_\mu \right) \nonumber \\
	&- \int\md^4x\,\sqrt{-g}\,\lambda \left(X + 1\right). \label{eq:Lagrangian_before_GW}
	\end{align}
	
	\section{Constraints from GW170817}
	\label{sec:Constraints_from_GW}
	As already mentioned, due to the detection of GW170817 \cite{TheLIGOScientific:2017qsa}, there are strict constraints on the propagation speed of gravitational waves, $c_t$, which has to be 
	equal to the speed of light, $c=1$, 
	up to very high accuracy, $\vert c_t^2/c^2-1\vert \le 5\times 10^{-16}$.
	In order to avoid any fine tuning, we simply require here that $c_t=c=1$. 
	Evaluating at linear order the tensor part of the EOM in conformal time, as
	\begin{align}
	c_1 h_{\alpha\beta}^{\prime\prime} + c_2 h_{\alpha\beta}^\prime + c_3 h_{\alpha\beta} + k^2 c_4 h_{\alpha \beta} =0,
	\end{align}
	where $h_{\alpha\beta}$ is the transverse and traceless metric perturbation and primes denote differentiation w.r.t. conformal time,
	the speed of gravitational waves can be calculated as
	\begin{equation}
	c_t^2 = \frac{c_4}{c_1}.
	\end{equation}
	The explicit expressions for the functions $c_i$ are given in the appendix. For the aforementioned Lagrangian Eq. \eqref{eq:Lagrangian_before_GW} we obtain for the propagation speed of 
	gravitational waves
	\begin{equation}
	c_t^2 = \frac{- 2 f(\varphi)}{-2 f(\varphi) - 2 a_1(\varphi) \frac{\varphi^{\prime 2}}{a^2} + b_2(\varphi) \left( 4 \frac{\h \varphi^{\prime 3}}{a^4} + 2 \frac{\varphi^{\prime ^2 \varphi^{\prime\prime}}}{a^4} 
	\right) + 6 b_3(\varphi) \frac{\h \varphi^{\prime 3}}{a^4}},
	\end{equation}
	where $\h=a^\prime/a$ is the Hubble parameter in conformal time. In \cite{Ezquiaga:2017ekz} it is discussed that the contributions to the propagation speed of the gravitational waves from the 
	terms of the second degree ($a_1$ and $a_2$) enter in a different part of the effective metric than the terms from third degree ($b_1, b_2$ and $b_3$). Further, it is not possible to tune the 
	functions between different parts of the effective metric in order to get a propagation speed of gravitational waves $c_t=1$. Using this fact, one easily realizes that the only natural way to get a 
	tensor speed equal to the speed of light is
	\begin{align}
	a_1(\varphi)=0,& \quad & b_3(\varphi) = - b_2(\varphi) \left(\frac{2}{3} + \frac{1}{3}\frac{\varphi^{\prime\prime}}{\h \varphi\prime}\right).
	\end{align}
	By using the background equation $\varphi^\prime = a$ one obtains the simplified condition $b_2=-b_3$, which is, however, completely background-dependent. In order to ensure that the speed of gravitational waves is equal to the speed of light in any background we have to restrict $b_2=b_3=0$. 
	Consequently, the action after the constraints due to the speed of gravitational waves simplifies to
	\begin{align}
	S &= \int\md^4x\,\sqrt{-g}\, \left( f_2 R + V + a_2 \left(\Box \varphi\right)^2  +  b_1 \left(\Box\varphi\right)^3 \right) - \int\md^4x\,\sqrt{-g}\,\lambda \left(X + 1\right)\,.
	\end{align}
	The presence of $a_2$ and $b_1$ alone without the counter terms would introduce the Ostrogradski instability. Therefore, we must set $a_2=b_1=0$ if we restrict ourselves to healthy seed actions (see the end of the section for a more detailed discussion). 
	
	Consequently, after imposing the constraints due to the gravitational waves and the conditions for a healthy seed action every higher-derivative term has to vanish, namely 
	\begin{equation}
	 a_1=a_2=b_1=b_2=b_3=0\,. \label{condition:healthy_mimetic}
	\end{equation}
	
	\noindent {\bf Alternative approach:} We may alternatively establish the above by first constraining the most general scalar-tensor theory from the GW170817 event, and followed by the effect of the mimetic constraint.
	The equation of motion for the mimetic scalar-tensor model in Eq. \eqref{a:m:dhost} can be written as 
	\begin{equation}
	E_{\mu\nu} + T_{\mu\nu} = - \left( E + T \right) \partial_\mu \varphi \partial_\nu \varphi.
	\end{equation}
	Since the metric perturbations for the tensor mode are traceless and transverse, the right-hand side does not contribute to the equation of motion for the tensor part at first order. Therefore, in general the conditions for the speed of the gravitational waves at linear order will be the same as in the seed theory, prior to the mimetic constraint. In principle, the background equation could cancel some terms, that contribute to the speed of the tensor mode, but this effect would of course be completely background dependent. In order to have valid constraints, independent of any background these ad-hoc solutions should be discarded. For that reason the conditions from the speed of the gravitational waves are the same for the seed theory and for the mimetic case. Therefore, if we want to start from the most general healthy scalar-tensor seed action after GW170817 we could directly start from the reduced DHOST theory discussed in \cite{Langlois:2017dyl,Crisostomi:2017lbg}. However, all the higher-derivative terms, which are left, have the form of $\theta_n$. Therefore, they do not contribute to the equation of motion in the mimetic matter theory. Hence, they can be neglected and one consequently obtains the final most general (healthy) mimetic scalar-tensor theory
	\begin{equation}
	\label{eq:Mimetic_scalar_tensor_after_GW}
	\int \md^4x\,\sqrt{-g}\,\left(f(\varphi) R + V(\varphi) \right) - \int \md^4x\,\sqrt{-g}\,\lambda \left(X+1\right).
	\end{equation}
	Therefore we come to the same conclusion previously mentioned in the Eq. \eqref{condition:healthy_mimetic}.
	
	In principle, one could suppose that there is no need for the degeneracy conditions of the DHOST model, since the mimetic constraint itself eliminates the fourth degree of freedom, due to the 
	higher-derivative terms \cite{Takahashi:2017pje}. So we could add terms as $(\Box \varphi)^2$, or in general functions as $f(\Box \varphi)$, which do not change the propagation speed of 
	gravitational waves. In fact, the term with $ (\Box \varphi)^2$ is widely discussed for the mimetic model, since it provides a non-vanishing scalar sound speed \cite{Chamseddine:2014vna}. On the 
	other hand, in the mimetic Horndeski model the sound speed for the scalar field vanishes, as discussed in \cite{Arroja:2017msd}. However, as discussed in the literature, mimetic theories with 
	this type of terms suffer of ghost or gradient instabilities \cite{Zheng:2017qfs,Firouzjahi:2017txv}.
	
	In the following we will restrict ourselves to the case $a_1=a_2=b_1=b_2=b_3=0$, which is viable in terms of the speed of gravitational waves and free from instabilities, as long 
	as $\lambda>0$ \cite{Chaichian:2014qba,MimeticHamiltonian}.

	\section{Poisson Equation for General Relativity}
	\label{sec:Poisson_Equation_for_General_Relativity}
	\subsection{Standard way}
	We study linear perturbations around a flat FLRW background. Considering only scalar perturbations in the Poisson gauge, and using conformal time $\eta$, we have
	\begin{equation}
	\md s^2 = - a^2 (\eta)\,(1 + 2 \Phi) \md \eta^2 + a^2(\eta)\, (1-2\Psi) \md x_i \md x^i.
	\end{equation}
	For the matter sector we consider a perfect fluid, $T_{\mu\nu}=(\rho + p)\, u_\mu u_\nu + p\, g_{\mu\nu}$. Due to the vanishing of the anisotropic stress, the traceless spatial 
	components of the Einstein's equations yield $\Phi=\Psi$.
	
	To calculate the Poisson equation in GR there is a standard way (see, e.g. \cite{Bardeen:1980kt,Mukhanov:1990me,Kodama:1985bj}). 
	First, one considers first-order perturbations of the time-time component of Einstein's equations $G_{00}^{(1)} = T_{00}^{(1)}$
	\begin{align}
	\label{eq:Time_Time_GR}
	2 \nabla^2 \Phi - 6 \h \,(\Phi^\prime + \h \Phi) &= a^2 \delta \rho.
	\end{align}
	As a next step the linearly perturbed time-space component of Einstein's equations are used, $G_{0i} = T_{0i}$, which gives
	\begin{align}
	\partial_i (\Phi^\prime + \h \Phi) &= -\frac{1}{2}a^2(\rho_0 + p_0) \partial_i v.
	\label{eq:First_Order_Space_Time_GR}
	\end{align}
	Integrating the previous equation and inserting it in the time-time component Eq. \eqref{eq:Time_Time_GR} we obtain the Poisson equation
	\begin{equation}
	\nabla^2 \Phi = 4\pi G\, a^2\rho_0 \Delta,
	\end{equation}
	with the gauge-independent density perturbation $\rho_0\Delta=\rho_0\delta-3\h(\rho_0+p_0)v$.
	
	\subsection{Alternative way}
	Later, we will show that the standard procedure cannot be applied in mimetic gravity theories. Therefore, we present here an alternative way 
	to derive the Poisson equation for the standard GR case.
	
	Integrating Eq. \eqref{eq:First_Order_Space_Time_GR} one obtains the velocity of the fluid as
	\begin{align}
	v &= - \frac{2}{a^2(\rho_0 + p_0)}(\Phi^\prime + \h \Phi) \nonumber \\
	&= - \frac{2}{a^3(\rho_0 + p_0)}(a\Phi)^\prime.
	\end{align}
	By linearly perturbing the energy continuity equation $\nabla_{\mu}T^{\mu 0}=0$, one gets an equation for the density fluctuation
	\begin{align}
	\label{eq:Energy_Continuity_Equation_GR}
	\delta\rho^\prime + 3\h (\delta\rho + \delta p)-3(\rho_0+p_0)\Psi^\prime+(\rho_0+p_0) \nabla^2 v =0.
	\end{align}
	Further, only non-relativistic matter (representing e.g. the baryonic component) is considered, $p_0\simeq \delta p \simeq 0$. 
	Using $\Phi=\Psi$ this yields in Fourier space
	\begin{align}
	\frac{1}{a^3}(a^3\delta\rho)^\prime - 3\rho_0 \Phi^\prime - \rho_0 k^2 v =0.
	\end{align}
	On small scales $k \gg \h $ one can neglect the second term compared to the third one in Eq. \eqref{eq:Energy_Continuity_Equation_GR}, since
	\begin{align}
	\rho_0 k^2 v &\sim \frac{k^2}{a^2} (\h \Phi + \Phi^\prime) \sim \frac{k^2}{a^2} \Phi^\prime, \\
	3 \rho_0 \Phi^\prime &\sim \frac{\h^2}{a^2}\Phi^\prime.
	\end{align}
	Finally, we can insert the expression for the velocity in the continuity equation to obtain the Poisson equation on small scales:
	\begin{align}
	\frac{1}{a^3}(a^3\delta \rho)^\prime &= - \frac{2 k^2}{a^3}(a\Phi)^\prime,
	\end{align}
	which yields, after performing the integration reinserting the reduced Planck mass $M_{\mathrm{pl}}^{-2}=8\pi G=1$, and going back to configuration space,
	\begin{align}
	\nabla^2 \Phi &= 4\pi G\,a^2 \delta\rho.
	\end{align}

	\subsection{Poisson equation for GR+CDM+PFDE}
	\label{sec:Poisson_equation_GR_CDM_PFDE}
	In this section the Poisson equation for GR plus cold dark matter and perfect fluid dark energy, with vanishing sound speed, is derived by using the aforementioned alternative 
	procedure. Further, the special case of a $\Lambda$CDM model is investigated separately. For a better comparison with the Poisson equation for mimetic matter our aim is to include all the 
	effects of the dark matter in an effective gravitational constant as felt by ordinary matter. It is important to note that, by making our choice even in the $\Lambda$CDM model, the effective 
	gravitational constant becomes time- and scale-dependent, due to the fact that we absorbed the dark matter contribution into the effective gravitational constant.
	
	\subsubsection*{PFDE background}
	The gravitational equation of motion can be written as
	\begin{equation}
	G_{\mu\nu} = T_{\mu\nu} + \tilde{T}_{\mu\nu} + \hat{T}_{\mu\nu},
	\end{equation}
	where $T^{\mu\nu}=\rho_{\mathrm{b}} u_{\mathrm{b}}^{\mu}u_{\mathrm{b}}^\nu$ is the energy-momentum tensor for the non-relativistic baryonic component. 
	Further, $\tilde T^{\mu\nu}=\rho_{\mathrm{dm}} u_{\mathrm{dm}}^\mu u_{\mathrm{dm}}^\nu$ is the energy-momentum tensor for the dark matter and $\hat{T}^{\mu\nu}=(\rho_{\mathrm{de}} + 
	p_{\mathrm{de}}) u_{\mathrm{de}}^\mu u_{\mathrm{de}}^\nu+  p_{\mathrm{de}} g^{\mu\nu}$ for the dark energy fluid.
	
	At the background level, the equations read
	\begin{align}
	\label{eq:Hubble_constraint_PFDE}
	\rho_{\mathrm{dm}} a^2 + \rho_{\mathrm{b}} a^2 +\rho_{\mathrm{de}} a^2&= 3 \h^2,  \\
	\h^2 + 2\h^\prime &=-a^2 p_{\mathrm{de}} \label{eq:Hubble_evolution_PFDE}, \\
	\rho_{l}^\prime + 3 \h (\rho_l + p_l) &= 0,
	\end{align}
	where $l \in \{\mathrm{dm},\mathrm{b},\mathrm{de}\}$. Since only non-relativistic baryons are considered $p_l  \approx 0$ for $l\in\{\mathrm{dm},\mathrm{b}\}$. At first order in the perturbations 
	the equations of motion for the space-time component of Einstein's equation and the continuity equation for the energy-momentum tensor are
	\begin{align}
	- \frac{2}{a}(a\Phi)^\prime &= a^2 \sum_{l} (\rho_l + p_l) v_l, \\
	\delta \rho_l^\prime + 3\h \delta \rho_l  &= 3(\rho_l +p_l) \Phi^\prime + (\rho_l+p_l) k^2 v_l, \\
	0&=\h  v_l + v_l^\prime +  \Phi,
	\label{eq:Momentum_Continuity_GR_CDM_PFDE}
	\end{align}
	where the last two equations are for $l\in\{\mathrm{dm},\mathrm{b}\}$. Using the momentum conservation we obtain
	\begin{equation}
	\label{eq:Metric_Potential_GR_CDM_PFDE}
	\Phi = - \frac{(a v_l)^\prime}{a}\quad \forall\, l\in\{\mathrm{dm},\mathrm{b}\}.
	\end{equation}
	From the space-time component of the metric equations we express the velocity of the baryons as
	\begin{equation}
	\label{eq:Velocity_Baryons_GR_CDM_PFDE}
	v_\mathrm{b}=-\frac{1}{a^3 \rho_{\mathrm{b}}}\left(2 (a\Phi)^\prime + a^3 \rho_{\mathrm{dm}} v_{\mathrm{dm}}+ a^3 (\rho_{\mathrm{de}} + p_{\mathrm{de}}) v_{\mathrm{de}}  \right).
	\end{equation}
	Using the energy conservation for the baryons one gets
	\begin{align}
	\frac{1}{a^3}\left(a^3 \delta\rho_{\mathrm{b}}\right)^\prime - 3 \rho_{\mathrm{b}} \Phi^\prime = k^2 \rho_{\mathrm{b}} v_{\mathrm{b}}, \nonumber \\
	\frac{1}{a^3}\left(a^3 \delta\rho_{\mathrm{b}}\right)^\prime \simeq k^2 \rho_{\mathrm{b}} v_{\mathrm{b}},
	\end{align}
	where in the second line the approximation of small scales $k\gg \h$ has been assumed, as before, by using Eq. \eqref{eq:Velocity_Baryons_GR_CDM_PFDE}.
	Inserting the expression for the velocity of the baryons yields
	\begin{align}
	\frac{1}{a^3}\left(a^3 \delta\rho_{\mathrm{b}}\right)^\prime &= -\frac{k^2}{a^3} \left(2 (a\Phi)^\prime + a^3 \rho_{\mathrm{dm}} v_{\mathrm{dm}}+ a^3 (\rho_{\mathrm{de}} + p_{\mathrm{de}}) 
	v_{\mathrm{de}}  \right), 
	\end{align}
	which yields, by using Eq. \eqref{eq:Momentum_Continuity_GR_CDM_PFDE} and performing the integration,
	\begin{align}
	\left(a^3 \delta\rho_{\mathrm{b}}\right)^\prime &= -k^2 \left(a^3 \rho_{\mathrm{dm}} v_{\mathrm{dm}}+ a^3 (\rho_{\mathrm{de}} + p_{\mathrm{de}}) v_{\mathrm{de}} - 2 (a 
	v_{\mathrm{dm}})^{\prime\prime}\right), \nonumber \\
	\delta \rho_{\mathrm{b}} &= -\frac{k^2}{a^3} \left[ \int \md \tilde{\eta}\,\left( a^3 \left(\rho_{\mathrm{dm}} v_{\mathrm{dm}}+ (\rho_{\mathrm{de}} + p_{\mathrm{de}}) v_{\mathrm{de}}\right)\right) - 2 (a 
	v_{\mathrm{dm}})^\prime\right].
	\label{eq:Density_Fluctuation_in_terms_of_field_fluctuations_PFDE}
	\end{align}
	Alternatively, one can obtain from the energy continuity equation for the baryons
	\begin{align}
	\left(a^3 \delta\rho_{\mathrm{b}}\right)^\prime &=-2k^2 (a\Phi)^\prime -k^2 a^3 \left( \rho_{\mathrm{dm}} v_{\mathrm{dm}}+  (\rho_{\mathrm{de}} + p_{\mathrm{de}}) v_{\mathrm{de}} \right),
	\end{align}
	which brings to
	\begin{align}
	-k^2 \Phi &= \frac{1}{2} a^2 \delta\rho_{\mathrm{b}} + \frac{k^2}{2a}\int\md \tilde{\eta}\, a^3\left( \rho_{\mathrm{dm}} v_{\mathrm{dm}}+  (\rho_{\mathrm{de}} + p_{\mathrm{de}}) v_{\mathrm{de}} 
	\right).
	\label{eq:Poisson_GR_CDM_PFDE_previous_step}
	\end{align}
	Using eqs. \eqref{eq:Density_Fluctuation_in_terms_of_field_fluctuations_PFDE} and \eqref{eq:Poisson_GR_CDM_PFDE_previous_step} the dark energy and dark matter contributions can be 'absorbed' into an effective gravitational constant
	\begin{align}
	-k^2 \Phi &= 4\pi G_{\mathrm{eff}}a^2\delta\rho_{\mathrm{b}},
	\end{align}
	where
	\begin{align}
	G_{\mathrm{eff}} = G \left(1+ \frac{\int \md \tilde{\eta}\, a^3 \left( \rho_{\mathrm{dm}} v_{\mathrm{dm}}+  (\rho_{\mathrm{de}} + p_{\mathrm{de}}) v_{\mathrm{de}} \right) }{ 2 (a v_{\mathrm{dm}})^\prime - \int \md \tilde{\eta}\, a^3 \left(\rho_{\mathrm{dm}} v_{\mathrm{dm}}+ (\rho_{\mathrm{de}} + p_{\mathrm{de}}) v_{\mathrm{de}}\right)}\right).
	\end{align}
	Assuming PFDE with vanishing sound speed the dark energy has a dust-like velocity (see, e.g. \cite{Arroja:2017msd})
	\begin{align}
		\Phi = - \frac{(a v_{\mathrm{de}})^\prime}{a}.
	\end{align}
	Using Eq. \eqref{eq:Metric_Potential_GR_CDM_PFDE} the velocities of the different fluids just differ by an integration constant $v_{\mathrm{dm}} = v_{\mathrm{de}} + \mathrm{const.}$
	Assuming, further, the initial condition $v_{\mathrm{de}}=v_{\mathrm{dm}}$ the effective gravitational constant simplifies to
	\begin{equation}
	G_{\mathrm{eff}} = G \left(1+ \frac{\int \md \tilde{\eta}\, a^3  v_{\mathrm{dm}} \left( \rho_{\mathrm{dm}}+\rho_{\mathrm{de}} + p_{\mathrm{de}} \right) }{ 2 (a v_{\mathrm{dm}})^\prime - \int \md 
	\tilde{\eta}\, a^3 v_{\mathrm{dm}} \left(\rho_{\mathrm{dm}} + \rho_{\mathrm{de}} + p_{\mathrm{de}} \right)}\right).
	\label{eq:Effective_Newtonian_Constant_PFDE_GR}
	\end{equation}
	
	\subsubsection*{$\mathbf{\Lambda}$CDM background}
	For a $\Lambda$CDM model the EOM are
	\begin{align}
	G_{\mu\nu} + \Lambda g_{\mu\nu} = T_{\mu\nu} + \tilde{T}_{\mu\nu},
	\end{align}
	which is a special case of the previous model with the additional condition $\Lambda=\rho_{\mathrm{de}} =-p_{\mathrm{de}}=\mathrm{const.}$ Using the previous calculation, the Poisson 
	equation can be written as
	\begin{equation}
	-k^2 \Phi = 4\pi G_{\mathrm{eff}}a^2\delta\rho_{\mathrm{b}},
	\end{equation}
	where
	\begin{align}
	G_{\mathrm{eff}} &= G \left(1+ \frac{\int \md \tilde{\eta}\, a^3  v_{\mathrm{dm}} \rho_{\mathrm{dm}}}{ 2 (a v_{\mathrm{dm}})^\prime - \int \md \tilde{\eta}\, a^3 v_{\mathrm{dm}} \rho_{\mathrm{dm}} }
	\right), \nonumber \\
	&= G \left(1+  \frac{C\int \md \tilde{\eta}\,  v_{\mathrm{dm}} }{ 2 (a v_{\mathrm{dm}})^\prime - C \int \md \tilde{\eta}\, v_{\mathrm{dm}}  }\right),
	\label{eq:Effective_Newtonian_Constant_Lambda_GR}
	\end{align}
	where in the second line we accounted for the fact that, at the background level, $\rho_{\mathrm{dm}}a^3=C=\mathrm{const.}$

	\section{Mimetic gravity}
	\label{sec:Mimetic_Gravity}
	The EOM of the mimetic scalar-tensor theory can be written as \cite{Arroja:2015yvd}
	\begin{align}
	& g^{\mu\nu}\partial_\mu \varphi \partial_\nu \varphi -1 = 0, \\
	& E_{\mu \nu} + T_{\mu\nu} = - (E+T) \partial_\mu \varphi \partial_\nu \varphi, \\
	& \Omega_\varphi + \partial_\mu \left(\sqrt{-g} (E+T) g^{\mu\nu}\partial_\nu \varphi\right)= 0, \\ 
	& \nabla_\mu T^{\mu\nu}=0. 
	\end{align}
	The equations are, however, not completely independent. The Klein-Gordon equation and the time-time component and the trace part of the spatial component of the Einstein's equation can be 
	derived 
	from the other equations (see the discussion in \cite{Arroja:2015yvd}).
	
	\subsection{Standard way}
	For the standard way of obtaining the Poisson equation we consider the time-time component at linear order, as in the previous section 
	\begin{align}
	E_{00}^{(1)} + T_{00}^{(1)} =- (E^{(1)}+ T^{(1)}) \varphi^{\prime 2} - 2 (E^{(0)} + T^{(0)}) \varphi^\prime \delta\varphi^\prime.
	\end{align}
	Using the first-order trace $E^{(1)}$
	\begin{align}
	E^{(1)} = -a^{-2}\left(E_{00}^{(1)}-\delta^{ij}E_{ij}^{(1)}-2\Phi E_{00}^{(0)}+2\Psi \delta^{ij}E_{ij}^{(0)}\right), \nonumber
	\end{align}
	and the mimetic constraint $\varphi^\prime = a$, we obtain for the time-time component
	\begin{align}
	E_{00}^{(1)} + T_{00}^{(1)} &= E_{00}^{(1)} - a^2 T^{(1)}- 2 a\left(E^{(0)}+T^{(0)}\right) \delta\varphi^\prime - \delta^{ij} E_{ij}^{(1)} - 2\Phi E_{00}^{(0)} + 2 \Psi \delta^{ij}E_{ij}^{(0)}.
	\end{align}
	We see that the zero-zero component $E_{00}^{(1)}$ is canceled. Therefore, the standard way of calculating the Poisson equation cannot be followed here.
	
	\subsection{Alternative way for the mimetic gravity }
	At the background level the independent equations for the mimetic gravity from Eq. \eqref{eq:Mimetic_scalar_tensor_after_GW} are 
	\begin{align}
	& \varphi^\prime = \pm a, \\
	& 3 \h \rho_\mathrm{b} + \rho_\mathrm{b}^\prime =0, \\
	& f \left(2 \h^2 + 4 \h^\prime \right)+ 4 a \h f_\varphi + 2 a^2 f_{\varphi\varphi}+ a^2 V  = 0.
	\label{eq:mimetic_matter_background_equation_general_case}
	\end{align}
	In the following, we will assume that $\varphi^\prime = + a$. At first order in the perturbations the metric constraint, the traceless (i,j) and the (0,i) component of the Einstein's equation and the 
	energy and momentum continuity equation are
	\begin{align}
	\label{eq:Horndeski_mimetic_Matter_1st_order_1}
	&\Phi = \varphi^{-1}  \delta \varphi^\prime =  a^{-1} \delta \varphi^\prime, \\
	\label{eq:Horndeski_mimetic_Matter_1st_order_2}
	&  f \Phi -  f \Psi + f_\varphi \delta\varphi =0, \\
	&  2 f_\varphi \delta\varphi^\prime + B\, \delta\varphi -4 f \Psi^\prime - \left(4 f \h +2 a f_\varphi \right) \Phi - a^2 \rho_\mathrm{b} v_\mathrm{b} = 0, \label{eq:Velocity_Mimetic_0i_component}\\
	&\delta \rho_\mathrm{b}^\prime + 3\h \delta \rho_\mathrm{b} - 3 \rho_\mathrm{b} \Psi^\prime -\rho_\mathrm{b} k^2 v_\mathrm{b} = 0, \label{eq:Energy_Conservation_Mimetic}\\
	& (\rho_\mathrm{b} v_\mathrm{b})^\prime + 4\h \rho_\mathrm{b} v_\mathrm{b} + \rho_\mathrm{b} \Phi = 0,
	\end{align}
	where
	\begin{equation}
	B = (-2 f_\varphi \h + 2 a f_{\varphi\varphi} +a(E^{(0)}+T^{(0)})) = 12 f \frac{\h^2+ \h^\prime}{a} + 16 \h f_\varphi + a \left(4 V + 8 f_{\varphi\varphi} - \rho_{\mathrm{b}}\right).
	\label{eq:Mimetic_B}
	\end{equation}
	Before obtaining the Poisson equation it is also useful to consider the ``gravitational slip". From \eqref{eq:Horndeski_mimetic_Matter_1st_order_1} and 
	\eqref{eq:Horndeski_mimetic_Matter_1st_order_2} one can express the potential $\Psi$ in terms of the field fluctuations $\delta\varphi$
	\begin{equation}
	\Psi =  \Phi + \frac{f_\varphi}{f} \delta \varphi =  \frac{\delta\varphi^\prime}{\varphi^\prime} + \frac{f_\varphi}{f} \delta \varphi.
	\label{eq:Psi_in_terms_delta_phi}
	\end{equation}
	This can be used to calculate the gravitational slip $\gamma$ as
	\begin{align}
	\frac{\Psi}{\Phi} &= \frac{\left( \frac{\delta\varphi^\prime}{\varphi^\prime} + \frac{f_\varphi}{f} \delta \varphi\right)}{\frac{\delta\varphi^\prime}{\varphi^\prime}}  =  \left(1 + \frac{f_\varphi}{f} \frac{\delta\varphi}{\delta\varphi^\prime} \varphi^\prime\right), \nonumber \\
	&= 1 + \frac{\ln\left( f\right)^\prime}{\ln\left(\delta \varphi \right)^\prime}=\gamma(\eta,\delta\varphi,\delta\varphi^\prime)
	\label{eq:Horndeski_mimetic_matter_gravitational_slip}
	\end{align}
	The gravitational slip depends on the scale $k$ (due to the fluctuations of the field) and on time. One can see that the gravitational slip vanishes, {\it i.e.} $\gamma = 1$, if and only if the function 
	$f(\varphi)=\mathrm{const}$, which is the case for a minimally coupled model. Note that the quasi-static approximation, commonly used to calculate the gravitational slip in modified gravity models 
	(see, e.g. \cite{Pogosian:2016pwr,Silvestri:2013ne}), fails, due to the mimetic constraint $\Phi= \delta\varphi^\prime/\varphi^\prime$.
		
	The Poisson equation can now be calculated using the same steps as in the alternative way for GR. From Eq. \eqref{eq:Velocity_Mimetic_0i_component} the velocity of the baryons can be 
	expressed as
	\begin{align}
	v_\mathrm{b} &= \frac{1}{a^2 \rho_\mathrm{b}}\left(- 4 f \Psi^\prime + 2 f_\varphi \delta\varphi^\prime + B\,\delta\varphi - \left(4 f \h +2 a f_\varphi \right) \Phi\right) \nonumber \\
	&= \frac{1}{a^2 \rho_\mathrm{b}}\left(- 4 f\Psi^\prime - 4 f \h \Phi + B\,\delta\varphi \right),
	\label{eq:Horndeski_mimetic_Matter_velocity_1}
	\end{align}
	where in the second line we used $\Phi = \delta \varphi^\prime / a$. From Eq. \eqref{eq:Horndeski_mimetic_Matter_1st_order_2} we obtain
	\begin{align}
	\Phi =  \Psi - \frac{f_\varphi}{f} \delta \varphi.
	\end{align}
	This equation can be inserted into Eq. \eqref{eq:Horndeski_mimetic_Matter_velocity_1}.
	\begin{align}
	v_\mathrm{b} & = \frac{1}{a^2 \rho_\mathrm{b}}\left(- 4 f \Psi^\prime - 4 f \h \Psi + \left(B+ 4 f_\varphi \h \right)\,\delta\varphi \right) \nonumber \\
	& = \frac{1}{a^2 \rho_\mathrm{b}}\left( - 4 f \frac{(a \Psi)^\prime}{a} + \left(B+ 4 f_\varphi \h \right)\,\delta\varphi \right).
	\end{align}
	As a next step the energy continuity equation is considered, using the same approximations as in the GR case, namely $\rho k^2 v \gg \rho \Psi^\prime$, considering the regime $k \gg \h$.
	\begin{align}
	\delta \rho_\mathrm{b}^\prime + 3\h \delta \rho &= k^2 \rho_\mathrm{b} v_\mathrm{b}, \nonumber \\
	\frac{1}{a^3}(a^3\delta\rho_\mathrm{b})^\prime &= \frac{k^2}{a^3}\left(- 4 f (a \Psi)^\prime+a(B+ 4 f_\varphi \h ) \delta \varphi\right).
	\label{eq:Horndeski_mimetic_matter_energy_conservation_1}
	\end{align}
	The density fluctuation $\delta\rho_\mathrm{b}$ can be expressed in terms of the perturbations of the field $\delta \varphi$ and its derivatives. Using Eq. \eqref{eq:Psi_in_terms_delta_phi} and 
	performing the integration, one finds
	\begin{align}
	\delta \rho_\mathrm{b} &= -\frac{k^2}{a^3} \Delta(\eta,\delta\varphi,\delta\varphi^\prime,\delta\varphi^{\prime\prime}),
	\label{eq:delta_rho_Mimetic}
	\end{align}
	where we defined
	\begin{align}
	\Delta(\eta,\delta\varphi,\delta\varphi^\prime,\delta\varphi^{\prime\prime}) &= \int\limits^{\eta}\md \tilde{\eta}\,\left[ 4 f \left(\delta\varphi^{\prime\prime} + \left(\frac{f_\varphi}{f}
	a\delta\varphi\right)^\prime\right)-a(B+ 4 f_\varphi \h)\delta\varphi\right].
	\end{align}
	Using the relation $f^\prime =f_\varphi \varphi^\prime=f_\varphi a$, we can rewrite the previous equation as
	\begin{align}
	\Delta &= \int\limits^{\eta}\md \tilde{\eta}\,\left[ 4 f \left( \delta \varphi^{\prime\prime} + \left(\frac{f^\prime}{f}\delta \varphi\right)^\prime\right)-\left(a B + 4 f^\prime \h \right) \delta\varphi \right] 
	\nonumber \\
	&= \int\limits^{\eta}\md \tilde{\eta}\, \left[4 f \delta \varphi^{\prime\prime}+ 4 f^\prime \delta \varphi^\prime + \left(4 f^{\prime\prime} -4 \frac{f^{\prime 2} }{f}-4 f^\prime \h - aB\right)\delta \varphi \right] 
	\nonumber \\
	&= 4 f \delta \varphi^\prime + \int\limits^{\eta}\md \tilde{\eta}\, \left( 4 f^{\prime\prime}-4 \frac{f^{\prime 2} }{f} -4 f^\prime \h - aB \right) \delta \varphi,
	\end{align}
	where in the last step we have used the fact that $4 f \delta \varphi^{\prime\prime} + 4 f^\prime \delta\varphi^\prime = 4 \left(f \delta \varphi^\prime\right)^\prime$. Alternatively, with the replacement   
	$f^\prime = f_\varphi a$, we can finally write
	\begin{align}
	\Delta = 4 f \delta \varphi^\prime +\int\limits^{\eta}\md \tilde{\eta}\,\left[4 f_{\varphi\varphi}a^2- 4 a^2 \frac{f_\varphi^2}{f} - aB\right]\delta \varphi.
	\end{align}
	Going back to the energy continuity equation \eqref{eq:Energy_Conservation_Mimetic}, we notice that it can be written as
	\begin{align}
	\frac{1}{4 f} (a^3 \delta\rho_\mathrm{b})^\prime &=-k^2 (a\Psi)^\prime+ \frac{a k^2}{4 f} (B+ 4 f_\varphi \h)\delta \varphi,
	\end{align}
	from which, by integration, it follows that
	\begin{align}
	\frac{1}{4 f}  a^3 \delta \rho_\mathrm{b} + \int\limits^{\eta}\md \tilde{\eta}\,\frac{f_\varphi}{4 f^2} a^4\delta \rho_\mathrm{b} &= - k^2 a \Psi + k^2 \int\limits^{\eta}\md \tilde{\eta}\,\frac{a}{4 f} (B+ 4 
	f_\varphi \h)\delta \varphi. 
	\end{align}
	By using Eq. \eqref{eq:delta_rho_Mimetic} we obtain
	\begin{align}
	k^2 \Psi &= - \frac{1}{4 f}a^2 \delta\rho_\mathrm{b}  + \frac{k^2}{a } \int\limits^{\eta}\md \tilde{\eta}\,\frac{a f_\varphi}{4 f^2}\Delta+\frac{k^2}{a }\int\limits^{\eta}\md \tilde{\eta}\, \frac{a}{4 f} (B+ 4 
	f_\varphi \h)\delta \varphi.
	\end{align}
	 Finally, by using Eq. \eqref{eq:Psi_in_terms_delta_phi} the Poisson equation can be expressed as
	\begin{align}
	-k^2 \Phi = \frac{1}{4 f}a^2\delta \rho_\mathrm{b} + k^2\frac{f_\varphi}{f}\delta\varphi - \frac{k^2}{a}\int\limits^{\eta}\md \tilde{\eta}\,\left[ \frac{a}{4 f} (B+ 4 f_\varphi \h)\delta \varphi +\frac{a f_\varphi}
	{4 f^2} \Delta \right].
	\end{align}
	One can see that the field fluctuations act as a source for the Poisson equation and the effective gravitational constant becomes time-dependent.
	
	It is also possible to rewrite the Poisson equation in a more compact form. Using Eq. \eqref{eq:delta_rho_Mimetic} and reinserting the Planck mass, the Poisson equation can be expressed as
	\begin{align}
	\label{eq:Poisson_equation_Mimetic_Gravity}
	-k^2 \Phi = 4\pi G_{\mathrm{eff}} a^2 \delta \rho_\mathrm{b},
	\end{align}
	with 
	\begin{align}
	G_{\mathrm{eff}}(\eta,\delta\varphi,\delta\varphi^\prime,\delta\varphi^{\prime\prime}) = G \left(\frac{1}{2 f} + 2\,  \frac{\int\limits^{\eta}\md \tilde{\eta}\,\left[ \frac{a}{4 f} (B+ 4 f_\varphi \h)\delta \varphi 
	+\frac{a f_\varphi}{4 f^2} \Delta \right] -\frac{a f_\varphi}{f}\delta\varphi }{\Delta}\right).
	\label{eq:G_effective_Mimetic_previous}
	\end{align}
	Every influence of the field fluctuations can be recast as a modification of the gravitational constant, which is now time- and scale-dependent, due to the fluctuations of the fields.
	
	To compare the effective gravitational constant with the results from \eqref{eq:Effective_Newtonian_Constant_PFDE_GR} or \eqref{eq:Effective_Newtonian_Constant_Lambda_GR} it is 
	convenient to rewrite \eqref{eq:G_effective_Mimetic_previous} in a different form. By partial integration one can simplify it to
	\begin{align}
	\frac{G_{\mathrm{eff}}}{G}&=\frac{1}{2 f} + 2 \frac{\int\limits^{\eta}\md \tilde{\eta}\, \left[\frac{1}{4f} \left(a B + 4 \h f^\prime \right)\delta \varphi + \frac{1}{4f} \Delta^\prime\right] - \frac{1}{4f}\Delta - 
	\frac{f^\prime}{f}\delta\varphi}{\Delta} \nonumber \\
	&=2 \frac{\int\limits^{\eta}\md \tilde{\eta}\,\left[\frac{f^\prime}{f}\delta\varphi^\prime + \delta \varphi^{\prime\prime} + \left(\frac{f^{\prime\prime}}{f}-\frac{f^{\prime 2}}{f^2}\right)\delta \varphi \right]-
	\frac{f^\prime}{f}\delta \varphi}{\Delta}\nonumber \\
	&= 2 \frac{\int\limits^{\eta}\md \tilde{\eta}\,\left[\left(\frac{f^\prime}{f}\delta \varphi\right)^\prime + \delta \varphi^{\prime\prime}\right]-\frac{f^\prime}{f}\delta \varphi}{\Delta}= \frac{2 \delta 
	\varphi^\prime}{\Delta}.
	\end{align}
	We can finally conclude
	\begin{align}
	\label{eq:Effective_Newtonian_Constant_mimetic}
	G_{\mathrm{eff}} = \left(1 + \frac{\left(2-4f\right)\delta \varphi^\prime - \int\limits^{\eta}\md \tilde{\eta}\, \left[ 4 f^{\prime\prime}-4 \frac{f^{\prime 2} }{f} -4 f^\prime \h - aB \right]\delta \varphi}{4 f \delta 
	\varphi^\prime +  \int\limits^{\eta}\md \tilde{\eta}\, \left[ 4 f^{\prime\prime}-4 \frac{f^{\prime 2} }{f} -4 f^\prime \h - aB \right] \delta \varphi }\right)G .
	\end{align}
	
	\subsection{Minimally coupled mimetic gravity}
	We start by considering minimally coupled models $f=1/2$ \cite{Arroja:2015yvd,Arroja:2017msd}. As outlined before, the gravitational slip for this simplified model vanishes, $\gamma=1$. The effective gravitational constant can be further simplified
	\begin{align}
	\label{eq:Minimally_coupled_G_eff}
	G_{\mathrm{eff}}= G\left(1 + \frac{\int\limits^{\eta}\md \tilde{\eta}\,a B \delta\varphi }{2\delta\varphi^{\prime}-\int\limits^{\eta}\md \tilde{\eta}\,a B \delta\varphi }\right).
	\end{align}
	Before continuing the discussion of the Poisson equation it is useful to consider the Klein-Gordon equation of the scalar field $\varphi$
	\begin{align}
	\Omega_\varphi + \partial_\mu \left(\sqrt{-g}(E+T) g^{\mu\nu}\partial_\nu \varphi\right)= 0.
	\end{align}
	At the background level, it simplifies to
	\begin{align}
	(a^3 (E^{(0)}+T^{(0)}))^\prime =& a^4 V_\varphi =  a^3 V^\prime. 
	\end{align}
	By using \eqref{eq:Mimetic_B}, it can be written as 
	\begin{align}
	\nonumber \\
	(a^2 B)^\prime =& a^3 V^\prime,
	\end{align}
	with
	\begin{align}
	a^2 B = 6 a \left(\h^2+ \h^\prime\right)  + a^3 \left(4 V  - \rho_{\mathrm{b}}\right).
	\end{align}
	The background equation for the minimally coupled mimetic gravity model with non-relativistic baryons, $E_{ij}^{(0)}=0$, can be written as
	\begin{align}
	\label{eq:cubic_mimetic_Horndeski_background}
	V a^2 + 2 \h^\prime + \h^2 = 0.
	\end{align}
	
	\subsubsection*{PFDE background}
	Assuming a background with CDM and PFDE, one has, from \eqref{eq:Hubble_constraint_PFDE} and \eqref{eq:Hubble_evolution_PFDE},
	\begin{align}
	&\rho_{\mathrm{dm}} a^2 + \rho_{\mathrm{b}} a^2 +\rho_{\mathrm{de}} a^2= 3 \h^2, \label{eq:background_1}  \\
	& 2\h^\prime  + \h^2 +a^2 p_{\mathrm{de}}=0 \label{eq:background_2}.
	\end{align}
	This leads to the identification for the function $V=p_{\mathrm{de}}$ by \eqref{eq:cubic_mimetic_Horndeski_background} \footnote{Let us stress here that, due to our sign convention,
	 $V(\varphi)$ is negative and equal to the pressure.}. Moreover, by using eqs. \eqref{eq:background_1} and 
	\eqref{eq:background_2},  $a^2B$ can be identified as
	\begin{equation}
	a^2B = a^3(\rho_{\mathrm{dm}}+\rho_{\mathrm{de}}+p_{\mathrm{de}}).
	\end{equation}
	As a consistency check, the Klein Gordon equation has to be fulfilled
	\begin{align}
	a^3 V^\prime&= \left(a^2 B\right)^\prime \nonumber \\
	&=\left(a^3 (\rho_{\mathrm{dm}}+\rho_{\mathrm{de}}+p_{\mathrm{de}})\right)^\prime \nonumber \\
	&=3a^3 \h (\rho_{\mathrm{de}}+p_{\mathrm{de}}) + a^3(\rho_{\mathrm{de}}+p_{\mathrm{de}})^\prime \nonumber \\
	&= a^3 p_{\mathrm{de}}^\prime,
	\end{align}
	where the continuity equations for the dark matter and the dark energy are used. One can see that the Klein Gordon equation is consistent with the previous identifications. Further, by using Eq. 
	\eqref{eq:Minimally_coupled_G_eff} the effective gravitational constant can be rewritten, in the case of a PFDE background, as 
	\begin{align}
	G_{\mathrm{eff}}&= G\left(1 + \frac{\int \md \tilde{\eta}\, a^3 (\rho_{\mathrm{dm}}+\rho_{\mathrm{de}}+p_{\mathrm{de}}) \frac{\delta \varphi}{a}}{2 \delta\varphi^\prime - \int \md \tilde{\eta}\, a^3 
	(\rho_{\mathrm{dm}}+\rho_{\mathrm{de}}+p_{\mathrm{de}}) \frac{\delta \varphi}{a} }\right).
	\label{eq:Minimally_coupled_G_eff_PFDE}
	\end{align}
	Using the natural identification $v_{\mathrm{dm}}=-\delta\varphi/\varphi^\prime$ as in \cite{Arroja:2017msd} and comparing Eq. \eqref{eq:Minimally_coupled_G_eff_PFDE} with 
	\eqref{eq:Effective_Newtonian_Constant_PFDE_GR} we can see that the Poisson equation for the minimally coupled mimetic gravity model with a PFDE background is equivalent to the Poisson 
	equation of standard GR with the same background.
	This is nicely in full agreement with the previous results of \cite{Arroja:2015yvd} and \cite{Arroja:2017msd}. 
	\medskip
	
	The necessary conditions and identifications are summarised here as follows:
	\begin{itemize}
		\item dust-like velocity of the dark energy fluid (hence PFDE with $c_s^{(\mathrm{de})}=0$)
		\item initial condition $v_{\mathrm{dm}}=v_{\mathrm{de}}$
		\item Identifications: $V=p_{\mathrm{de}}$, $a^2B=a^3(\rho_{\mathrm{dm}}+\rho_{\mathrm{de}}+p_{\mathrm{de}})$ and $v_{\mathrm{dm}}=-\delta\varphi/\varphi^\prime$
	\end{itemize}
	
	\subsubsection*{$\mathbf{\Lambda}$CDM background}
	As already mentioned, the $\Lambda$CDM background is a special case of PFDE. For the function $V$, using the background equation, it yields the condition $V=\mathrm{const.}\equiv - 
	\Lambda $. For this case, the Klein-Gordon equation simplifies to 
	\begin{equation}
	\left(a^2 B\right)^\prime = \left( 6 a \left(\h^2+ \h^\prime\right)  + a^3 \left(-4 \Lambda  - \rho_{\mathrm{b}}\right)\right)^\prime = 0.
	\end{equation}
	The effective gravitational constant can be expressed as 
	\begin{equation}
	G_{\mathrm{eff}}= G\left(1 + \frac{C \int \md \tilde{\eta}\, \frac{\delta \varphi}{a}}{2 \delta\varphi^\prime - C \int \md \tilde{\eta}\,\frac{\delta \varphi}{a} }\right),
	\end{equation}
	where $a^2B=\mathrm{const.}\equiv C= a^3 \rho_{\mathrm{dm}}$. Using the identification $v_{\mathrm{dm}}=-\delta\varphi/\varphi^\prime$ again this is equivalent to the effective gravitational 
	constant for standard GR with a cosmological constant and cold dark matter (see Eq. \eqref{eq:Effective_Newtonian_Constant_Lambda_GR}).
	
	\subsection{The case of non-minimally coupled mimetic gravity}
	So far we have discussed the case of a minimally coupled model, which is equivalent to the mimetic cubic Horndeski model, using the two different backgrounds of PFDE and $\Lambda$CDM. 
	Therefore, it is interesting to consider a $\Lambda$CDM background in the case of a more general non-minimally coupled mimetic theory with $f(\varphi)\neq \mathrm{const.}$
	
	It is convenient to discuss the background equation in cosmic time. The metric equation in the FLRW background can be written as
	\begin{equation}
	2 f(3 H^2 + 2 \dot{H}) + V + 4 H  f_\varphi + 2  f_{\varphi\varphi} = 0.
	\end{equation}
	The mimetic constraint yields in cosmic time $\varphi = t$, where we have neglected the integration constant. In a $\Lambda$CDM background we can parametrise the scale-factor as 
	\begin{equation}
	a(t)=a_\star \sinh\left(C t\right)^{2/3}
	\end{equation}
	with $C= \sqrt{3 \Lambda/4}$ (see, e.g. \cite{Arroja:2015wpa}). Inserting it into the background equation yields a condition for the free functions $f$ and $V$
	\begin{align}
	2 \ddot f + \frac{8}{3} C \coth(C t) \dot f + \frac{8}{3} C^2 f = -V.
	\end{align}
	This is a second-order inhomogeneous differential equation for $f(t)$. Since $f$ and $V$ are free functions we can define for any function $f(t)$ an appropriate $V(t)$ in order to solve such a 
	differential equation. Therefore, it is in general not a problem to obtain a $\Lambda$CDM background. 
	
	Let us discuss now one single toy model to illustrate possible solutions. Defining $V(t)=-8/3 C^2 \ddot f- b$, where $b=\mathrm{const.}$, we obtain a differential equation of first order.
	\begin{equation}
	\frac{8}{3} C \coth(C t) \dot f(t) + \frac{8}{3} C^2 f(t) =b.
	\end{equation}
	This can be solved to yield
	\begin{align}
	f(t) = \frac{3 b}{8 C^2} + d \cosh(C t)^{-1} 
	\end{align}
	with integration constant $d$. In this way one could find solutions for the free functions with the desired behaviour. In this specific case, $f(t)$ starts from $3b/(8 C^2) +d$ and then approaches 
	exponentially the value of $3b/(8 C^2)$ which could be normalised to the current value of $1/2$ in units of $M_{\mathrm{pl}}^2=1$.
	\medskip
	
	For the case of a minimally coupled mimetic model $f=1/2$ we have shown that in the $\Lambda$CDM background the effective gravitational constant is equivalent to that of the standard 
	cosmological model consistently with previous results in \cite{Arroja:2017msd,Arroja:2015yvd}. Therefore, it is interesting to discuss if this is also the case for $f(\varphi)\neq \mathrm{const}$ in a 
	$\Lambda$CDM background. For a general function $f(\varphi)$ the effective gravitational constant is \eqref{eq:Effective_Newtonian_Constant_mimetic}
	\begin{equation}
	\label{eq:G_eff_non_minimally_coupled_mimetic}
	G_{\mathrm{eff}} = \left(1 + \frac{\left(2-4f\right)\delta \varphi^\prime - \int\limits^{\eta}\md \tilde{\eta}\, \left[ 4 f^{\prime\prime}-4 \frac{f^{\prime 2} }{f} -4 f^\prime \h - aB\right]\delta \varphi}{4 f \delta 
	\varphi^\prime +  \int\limits^{\eta}\md \tilde{\eta}\, \left[ 4 f^{\prime\prime}-4 \frac{f^{\prime 2} }{f} -4 f^\prime \h - aB\right] \delta \varphi }\right) G.
	\end{equation}
	As before, we can identify $\delta \varphi = - a v_{\mathrm{dm}}$. Comparing the previous equation with \eqref{eq:Effective_Newtonian_Constant_Lambda_GR}
	\begin{equation}
	G_{\mathrm{eff}}=  G \left(1+  \frac{C\int \md \tilde{\eta}\,  v_{\mathrm{dm}} }{ 2 (a v_{\mathrm{dm}})^\prime - C \int \md \tilde{\eta}\, v_{\mathrm{dm}}  }\right),
	\end{equation}
	we obtain the conditions $(2-4f)=0$ and $a \left( 4 f^{\prime\prime}-4 \frac{f^{\prime 2} }{f} -4 f^\prime \h - aB \right) =\mathrm{const.}$ to have equivalent solutions. Therefore, we can observe that 
	for any function $f(t)\neq 1/2$ it is not possible to obtain the same effective gravitational constant. The same conclusion can be obtained for a PFDE background model, comparing Eq. 
	\eqref{eq:G_eff_non_minimally_coupled_mimetic} with Eq. \eqref{eq:Effective_Newtonian_Constant_PFDE_GR}.
	
	The deviations of the Poisson equation in the non-minimally coupled mimetic gravity model from the standard $\Lambda$CDM model can be directly calculated by dividing the derived expression 
	of the effective gravitational constant for the mimetic gravity model in Eq. \eqref{eq:G_eff_non_minimally_coupled_mimetic} by the effective gravitational constant for the standard $\Lambda$CDM 
	model in Eq. \eqref{eq:Effective_Newtonian_Constant_Lambda_GR}
	\begin{align}
	\tilde G_{\mathrm{eff}} = \frac{G_{\mathrm{eff}}^{\mathrm{mim}}}{G_{\mathrm{eff}}^{\mathrm{\Lambda}}}\,G = \frac{2 \delta \varphi^\prime - C \int \md\tilde \eta\, \frac{\delta \varphi}{a}}{4 f \delta 
	\varphi^\prime +  \int\limits^{\eta}\md \tilde{\eta}\, \left[ 4 f^{\prime\prime}-4 \frac{f^{\prime 2} }{f} -4 f^\prime \h - aB\right] \delta \varphi }\,G.
	\end{align}
	Normalized to the gravitational constant  $\tilde G_{\mathrm{eff}}$ is the effective gravitational constant which is commonly used in the literature to express deviations from the standard 
	$\Lambda$CDM 
	model.
	
	\section{Conclusions}
	\label{sec:Conclusions}
	In the first part of the paper we have derived the most general mimetic scalar-tensor theory given by Eq. \eqref{eq:Mimetic_scalar_tensor_after_GW}, accounting for the constraint due to the 
	speed of the gravitational waves and assuming a healthy seed theory. In a subsequent paper \cite{MimeticHamiltonian} it will be shown that under the condition $\lambda\ge0$ this mimetic theory 
	is stable and it is not plagued by the presence of ghost instabilities.
	
	In the second part of this paper we have derived for this mimetic gravity model the Poisson equation \eqref{eq:Poisson_equation_Mimetic_Gravity} with the effective gravitational constant Eq. 
	\eqref{eq:Effective_Newtonian_Constant_mimetic}, by studying first order perturbations around a flat FLRW background. In the case of a minimally coupled mimetic gravity theory $f(\varphi)=1/2$ 
	we showed that the Poisson equation is equivalent to GR with CDM and PFDE, with vanishing sound speed. A similar result was obtained for the evolution equation of the metric perturbation in 
	\cite{Arroja:2017msd} for a mimetic cubic Horndeski model, which is equivalent to our model up to redefinitions. Restricting to a $\Lambda$CDM background implies that the minimally coupled 
	mimetic gravity model is not distinguishable from the standard $\Lambda$CDM model, up to linear order. The mimetic field mimics the CDM contribution while the dark energy is provided by a 
	cosmological constant $V(\varphi)=-\Lambda$. 
	In the last part we have shown that for the full non-minimally coupled mimetic gravity model $f(\varphi)\neq \mathrm{const.}$ the Poisson equation always differs from the standard 
	$\Lambda$CDM model even for a perfect $\Lambda$CDM background, which can be obtained by appropriately choosing $V(\varphi)$ and $f(\varphi)$. Further, in this case there is always a non-
	vanishing gravitational slip, as given by Eq. \eqref{eq:Horndeski_mimetic_matter_gravitational_slip}. 
	
	Altogether, the mimetic gravity theory with a non-minimally coupling provides an interesting modified gravity model which can be distinguished from the standard $\Lambda$CDM model at linear 
	order. Due to the mimetic constraint the non-minimally coupling of the scalar field has a different phenomenology, compared to standard scalar-tensor theories with an unconstrained scalar field. 
	Therefore, in a future work we plan to explicitly study the behaviour of the coupling in more detail and consider further consequences of it.
	
	\acknowledgments We warmly acknowledge Frederico Arroja and Dimitri Sorokin for useful discussions. Part of the computations are done using Mathematica\footnote{https://www.wolfram.com/
	mathematica/} with the algebra package xAct\footnote{http://www.xact.es/} and its contributed package xPand\footnote{http://www2.iap.fr/users/pitrou/xpand.htm} \cite{Pitrou:2013hga}. We acknowledge partial financial support by ASI Grant No. 2016-24-H.0. PK acknowledges financial support from ``Fondazione Ing. Aldo Gini''.

	\appendix
	\section{Explicit expression for the functions $c_i$}
	In this appendix we give the explicit expressions for the functions $c_i$, $i=1,...,4$ defined in the main text.
	\begin{align}
	c_1 =& -\frac{2}{a^4} \left( a^4 f + a_1 a^2 (\varphi^\prime)^2 -(\varphi^\prime)^2 \left(3 b_3 \h \varphi^\prime + b_2 \left(2\h \varphi^\prime + \varphi^{\prime\prime}\right)\right)\right) \\
	c_2 =& - \frac{2}{a^4} \Big( a^2 (2 a_1 \varphi^\prime \varphi^{\prime\prime} + \left(\varphi^\prime)^3 a_{1\varphi}\right)  - \varphi^\prime b_2 \left(- 4 \h^2 (\varphi^\prime)^2 + 2 \h^\prime 
	(\varphi^\prime)^2 + 4 \h \varphi^\prime \varphi^{\prime\prime} + 2 (\varphi^{\prime\prime})^2+ \varphi^\prime \varphi^{\prime\prime\prime}\right) \nonumber \\ & -(\varphi^{\prime})^2 b_3 \left(- 6 
	\h^2 \varphi^\prime + 3 \h^\prime \varphi^\prime + 9 \h \varphi^{\prime\prime}\right) - (\varphi^\prime)^3 \left( \varphi^{\prime\prime} b_{2\varphi} + \h \varphi^\prime \left(2 b_{2\varphi} + 3 
	b_{3\varphi}\right)\right)  \nonumber \\ &+ a^4 \left(2 f  \h + \varphi^\prime f_\varphi \right)\Big) \\
	c_3 =& \frac{2}{a^4} \Big( a^6 V - 24 b_2 \h^3 (\varphi^\prime)^3 - 8 b_3 \h^3 (\varphi^\prime)^3 + 16 b_2 \h \h^\prime (\varphi^\prime)^3 + 6 b_3 \h \h^\prime (\varphi^\prime)^3 + 24 b_2 \h^2 
	(\varphi^\prime)^2 \varphi^{\prime\prime}  \nonumber \\
	&+ 6 b_3\h^2 (\varphi^\prime)^2 \varphi^{\prime\prime} - 2 b_2 \h \varphi^\prime (\varphi^{\prime\prime})^2 + 3 b_3 \h \varphi^\prime (\varphi^{\prime\prime})^2 - b_3 (\varphi^{\prime\prime})^2 + 2 
	b_2 \varphi^\prime \varphi^{\prime\prime} \varphi^{\prime\prime\prime} - 2 b_1 \left( 2 \h \varphi^\prime + \varphi^{\prime\prime}\right) \nonumber \\
	&\times \left(8 \h^2 (\varphi^\prime)^2 - 6\h^\prime (\varphi^\prime)^2 - 4 \h \varphi^\prime \varphi^{\prime\prime} - (\varphi^{\prime\prime})^2 - 3\varphi^\prime \varphi^{\prime\prime\prime} \right) +  
	a^2 a_1 \Big(4 \h^2 (\varphi^{\prime})^2 - 2 \h^\prime (\varphi^\prime)^2 \nonumber \\ 
	&- 6 \h \varphi^\prime \varphi^{\prime\prime} + (\varphi^{\prime\prime})^2 \Big) + a^2 a_2 \left( 4 \h^2 (\varphi^{\prime})^2 - 4 \h^\prime (\varphi^\prime)^2 - 4 \h \varphi^\prime \varphi^{\prime\prime} 
	- (\varphi^{\prime\prime})^2 - 2\varphi^\prime \varphi^{\prime\prime\prime}\right) \nonumber \\
	&- 2 a^2 (\varphi^\prime)^2 \left(\varphi^{\prime\prime} a_{2\varphi} + \h \varphi^\prime \left(a_{1\varphi} + 2 a_{2\varphi}\right)\right) +  b_{1\varphi} \left(12 \h^2 (\varphi^\prime)^4 + 12 \h 
	(\varphi^\prime)^3 \varphi^{\prime\prime}  + 3 (\varphi^\prime)^2 (\varphi^{\prime\prime})^2\right)  \nonumber \\
	&+ b_{2\varphi} \left( 8 \h^2 (\varphi^\prime)^4 + (\varphi^\prime)^2 (\varphi^{\prime\prime})^2 \right) + 3 b_{3\varphi} \h^2 (\varphi^\prime)^4 + 2 a^4 \Big(f \left(\h^2+2\h^\prime\right)  + \h 
	\varphi^\prime f_\varphi \nonumber \\ 
	&+ \varphi^{\prime\prime} f_\varphi + (\varphi^\prime)^2 f_{\varphi\varphi}\Big) \Big)	\\
	c_4 =& - 2 f
	\end{align}
	
	\bibliography{bibliography_G_Newton}
	\bibliographystyle{JHEP}
	
\end{document}